# Computational Fluid Dynamics In GARUDA Grid Environment

*By Chandra Bhushan Roy, Dr Vikas Kumar*
*Centre for Development of Advance Computing ,*
*Computational Fluid Dynamics Team, Pune*
*E-mail:{chandrar , vikask}@cdac.in*

**Abstract**

### 1.Motivation

GARUDA Grid developed on NKN (National Knowledge Network) network by Centre for Development of Advanced Computing (C-DAC) hubs High Performance Computing (HPC) Clusters which are geographically separated all over India. C-DAC has been associated with development of HPC infrastructure since its establishment in year 1988. The Grid infrastructure provides a secure and efficient way of accessing heterogeneous resource . Enabling scientific applications on Grid has been researched for some time now. In this regard we have successfully enabled Computational Fluid Dynamics (CFD) application which can help CFD community as a whole in effective manner to carry out computational research which requires huge compuational resource beyond once in house capability. This work is part of current on-going project Grid GARUDA funded by Department of Information Technology.

### 2.Problem Statement

Computational Fluid Dynamics (CFD) is one of the driving force behind HPC evolution. This often requires huge computational resource time and again, which cannot be met all the time by existing physical resource at our workplace . Now to meet this often the researcher is required to be in queue and wait for the resource to get free. In this regard one loses reasonable time and money. The second problem is coupling of pre-processor , solver and post processor of CFD  tools makes it  difficult to prepare a working mathematical  model of physical  problem all the time. And some specific CFD tools require huge finance  to get working on cluster.

Both the problem requires attention for growth of CFD practice, because if resources are not available in time and tools end up being tightly coupled then this will lead the technology to be working  in only elite hands. Overcoming  the above limitation is must for researchers and also for industry. In this regard we put forward a working model of CFD over Grid in India.

### 3.Approch

In the current model a user can choose between highly flexible working model, starting with mesh generation , solver and post processor. Grid connects more then 20 research centres and  academic institutes spread over 8 cities, each one of it can be used from any place connected through NKN network. And jobs can be submitted in GUI and command line mode. Thus user has large resource under reach. This environment of Grid solves our first issue of resource shortage at any given time. Now we need to address our second problem i.e. CFD on Grid.

There can be two paths which would be followed by user. First, one can do the pre-processing of flow model on local desktop using application available commercial or open-source. And second if they don't have access to pre-processing application they can use the one available on Grid. Now the solver part is highly compute intensive which should be done on Grid using any one resource. Now after the results have been calculated, again we have two path which ever convenient should be followed i.e. either do the post processing on Grid resource or on local desktop.

In our current work we have used open source solver OpenFOAM and prepared a skeleton of this model . Here we would stress that this can be augmented with other solvers too. In this way user will have high flexibility in selecting suitable solver.

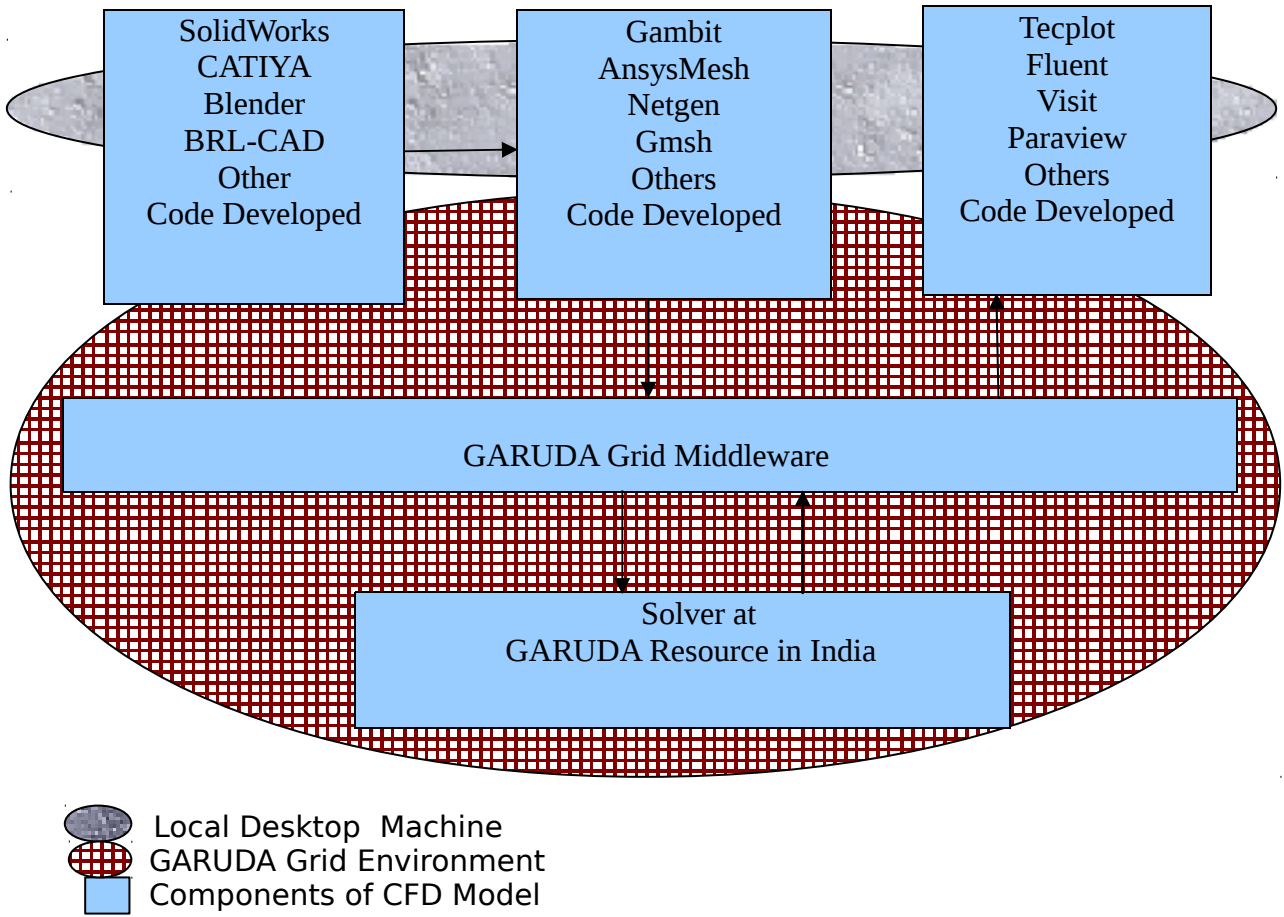

## 4.Result

Grid environment and the CFD model prepared will reduce cost of computation. If the current model is used it will also encourage usage of open source CFD code which again will reduce the commercial over head of CFD application.

## 5.Conclusion

Grid computing will become more efficient with this CFD model. At the same time CFD community would have a common environment where they can work and share knowledge. In future this model should also be applied in academic labs and industries.

## 6.CITATIONS

[1]Foster, Kesselman, The Anatomy of the Grid Enabling Scalable Virtual Organizations-2001
[2] http://www.garudaindia.in
[3]Xiaobo Yang, Mark Hayes "Application of Grid techniques in the CFD"Integrating CFD and Experiments in Aerodynamics Glasgow, UK, September 2003